\documentclass[prb,twocolumn,psfig,show pacs]{revtex4}
\usepackage{amsfonts}
\usepackage{amsmath}
\usepackage{graphicx}
\usepackage{bm}
\usepackage{amssymb}
\usepackage{times}
\usepackage{dcolumn}
\usepackage{cases}
\usepackage{txfonts}

\begin{document}

\title{Phase transitions and thermodynamics of the two-dimensional Ising model on a distorted Kagom\'{e} lattice}
\author{Wei Li$^{1}$, Shou-Shu Gong$^{1}$, Yang Zhao$^{1}$, Shi-Ju Ran$^{1}$, Song Gao$^{2}$, and Gang Su$^{1,{\ast }}$}
\affiliation{$^1$College of Physical Sciences, Graduate University of Chinese Academy of Sciences, P. O. Box 4588, Beijing 100049, People's Republic of China\\
$^2$College of Chemistry and Molecular Engineering, State Key Laboratory of Rare Earth Materials Chemistry and Applications, Peking University, Beijing 100871, People's Republic of China}

\begin{abstract}
The two-dimensional Ising model on a distorted Kagom\'{e} lattice is
studied by means of exact solutions and the tensor renormalisation
group (TRG) method. The zero-field phase diagrams are obtained,
where three phases such as ferromagnetic, ferrimagnetic and
paramagnetic phases, along with the second-order phase transitions,
have been identified. The TRG results are quite accurate and
reliable in comparison to the exact solutions. In a magnetic field,
the magnetization ($m$), susceptibility and specific heat are
studied by the TRG algorithm, where the $m=1/3$ plateaux are
observed in the magnetization curves for some couplings. The
experimental data of susceptibility for the complex
Co(N$_3$)$_2$(bpg)$\cdot$ DMF$_{4/3}$ are fitted with the TRG
results, giving the couplings of the complex $J=22K$ and $J'=33K$.
\end{abstract}

\maketitle

\section{Introduction}

\begin{figure}[tbp]
\includegraphics[angle=0,width=0.6\linewidth]{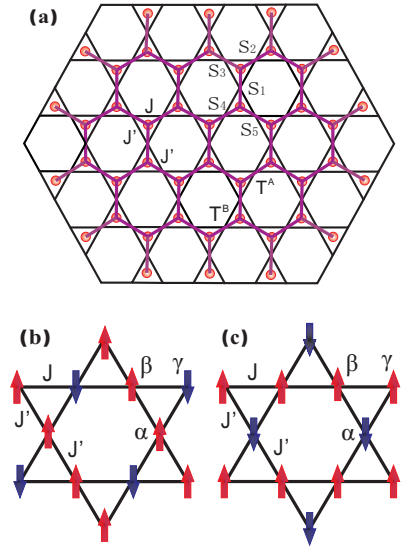}
\caption{(Color online) (a) The distorted Kagom\'{e} lattice, where
$J$ and $J'$ denote the different nearest neighbor couplings. The
dots in the center of small triangles and lines connecting them form
a tensor network presented by Eq. (\ref{formula-partition func}).
(b) and (c) show two degenerate spin configurations of the
ferrimagnetic structure, where up and down arrows represent the
spin-up and spin-down states, respectively.} \label{fig-DKL}
\end{figure}

Kagom\'{e} lattice, one of the most interesting frustrated spin
lattices, has attracted much attention both experimentally and
theoretically in recent years. The Heisenberg model on a Kagom\'{e}
lattice is likely a candidate for finding a spin liquid state in
two-dimensional (2D) spin systems. Some numerical works have
revealed that the system possesses a magnetic disordered ground
state.\cite{Lecheminant,Jiang0} Nevertheless, the nature of its
ground state is still an open question.\cite{Richter} Recently, a
number of spin systems, such as volborthite
Cu$_3$V$_2$O$_7$(OH)$_2\cdot$2H$_2$O (Refs.
\onlinecite{Wang,Schnyder,Yoshida,Yoshida2}),
[H$_3$N(CH$_2$)$_2$NH$_2$(CH$_2$)$_2$(NH$_3$]$_4$[Fe$^{\textrm{II}}_9$F$_{18}$(SO$_4$)$_6$]$\cdot$9H$_2$O
(Ref. \onlinecite{Behera}) and Co(N$_3$)$_2$(bpg)$\cdot$ DMF$_{4/3}$
(Ref. \onlinecite{Gao}), are found to form a Kagom\'{e} lattice with
distortions, where the structural distortions give rise to two
different exchange couplings $J$ and $J'$. Such a spatially bond
anisotropic spin lattice can be called a distorted Kagom\'{e} (DK)
lattice, as schematically depicted in Fig. \ref{fig-DKL}. A
distortion-induced magnetization step at small fields and the $1/3$
magnetization plateau on the DK lattice have been observed in
experiments. \cite{Yoshida2} In order to explain the experimental
results, the quantum and classical Heisenberg models on the DK
lattice are considered.\cite{Hida, Kaneko} In most cases, the
interactions between spins are usually of Heisenberg type on a DK
lattice \cite{Kaneko}. However, in some materials the spin-spin
couplings are anisotropic, and even of Ising-type in particular
situations. For instance, in complex Co(N$_3$)$_2$(bpg)$\cdot$
DMF$_{4/3}$, Co$^{+2}$ ions form a spin-1/2 DK lattice and might be
coupled by Ising-type interactions at low temperatures.\cite{Carlin}
Therefore, it should also be necessary to pay more attention on the
2D Ising model with the DK lattice, especially when an external
magnetic field is present, where the works in the literature are
still sparse.

In this article, we shall focus on the Ising model on the DK
lattice, where the thermodynamics and magnetic properties will be
carefully studied by means of exact solutions and a numerical
method. In the present model, the quantum fluctuations are
completely suppressed, and only the thermal fluctuations are
considered. It should be stressed that the present model for $h=0$
can be exactly solved, but for $h \neq 0$ the exact solution is not
available at present and the numerical method should be involved in.
For this reason, a recently developed tensor renormalization group
(TRG) method \cite{Levin,Jiang,Gu} is employed to investigate the
thermodynamic properties of the system. The cases with different
couplings $J$ and $J'$ and both with and without a magnetic field
$h$ will be discussed. The TRG results, being consistent with the
exact solutions, reveal that the system has ferromagnetic,
ferrimagnetic, and paramagnetic phases in the phase diagram, where
the paramagnetic phase can exist at $T=0$ in the strongly frustrated
region. Moreover, the magnetic order-disorder phase transitions
separating these three phases are disclosed. In ferrimagnetic and
paramagnetic phases, the $1/3$ magnetization plateaux are seen. A
magnetization step at an infinitesimal field is observed for the
paramagnetic phase at low temperature. The specific heat no longer
possesses a divergent peak once the external field is switched on,
implying the absence of phase transitions at $h\neq 0$. In addition,
we shall also make an attempt to fit the experimental data of
susceptibility for the complex Co(N$_3$)$_2$(bpg)$\cdot$ DMF$_{4/3}$
with the TRG results so as to estimate the exchange couplings in
this complex.

The other parts of this article are organized as follows. In Sec.
II, the exact solutions and phase diagrams are presented. The TRG
method is introduced in Sec. III. The specific heat without a
magnetic field is explored in Sec. IV. Magnetization,
susceptibility, and a comparison to the experimental data are shown
in Sec. V. Sec. VI contains the results of specific heat in an
external magnetic field. The summary and discussions are given
finally.

\section{Exact solutions in zero magnetic field}

The Hamiltonian of the system under interest has the form of
\begin{equation}
H=J' \sum_{<i\in \alpha, j\in \beta>}{S_i S_j}+ J' \sum_{<i\in \alpha, j\in\gamma>}{S_i S_j}+ J \sum_{<i\in \beta, j\in \gamma>}{S_i S_j}-h\sum_i{S_i},
\label{formular-Hamiltonian}
\end{equation}
where $S_i$ is an Ising spin with two discrete values $\pm 1$, and
the whole lattice can be divided into three sublattices labeled as
$\alpha$, $\beta$, and $\gamma$. The spin-spin coupling terms are
restricted to nearest neighbor sites, $J$ and $J'$ are different
nearest neighboring couplings as shown in Fig. \ref{fig-DKL}, where
$J(J')>0$ and $<0$ represent antiferromagnetic and ferromagnetic
couplings, respectively, $h$ is the uniform external magnetic field,
and $g \mu_B = 1$ is assumed.

\begin{figure}[tbp]
\includegraphics[angle=0,width=1.05\linewidth]{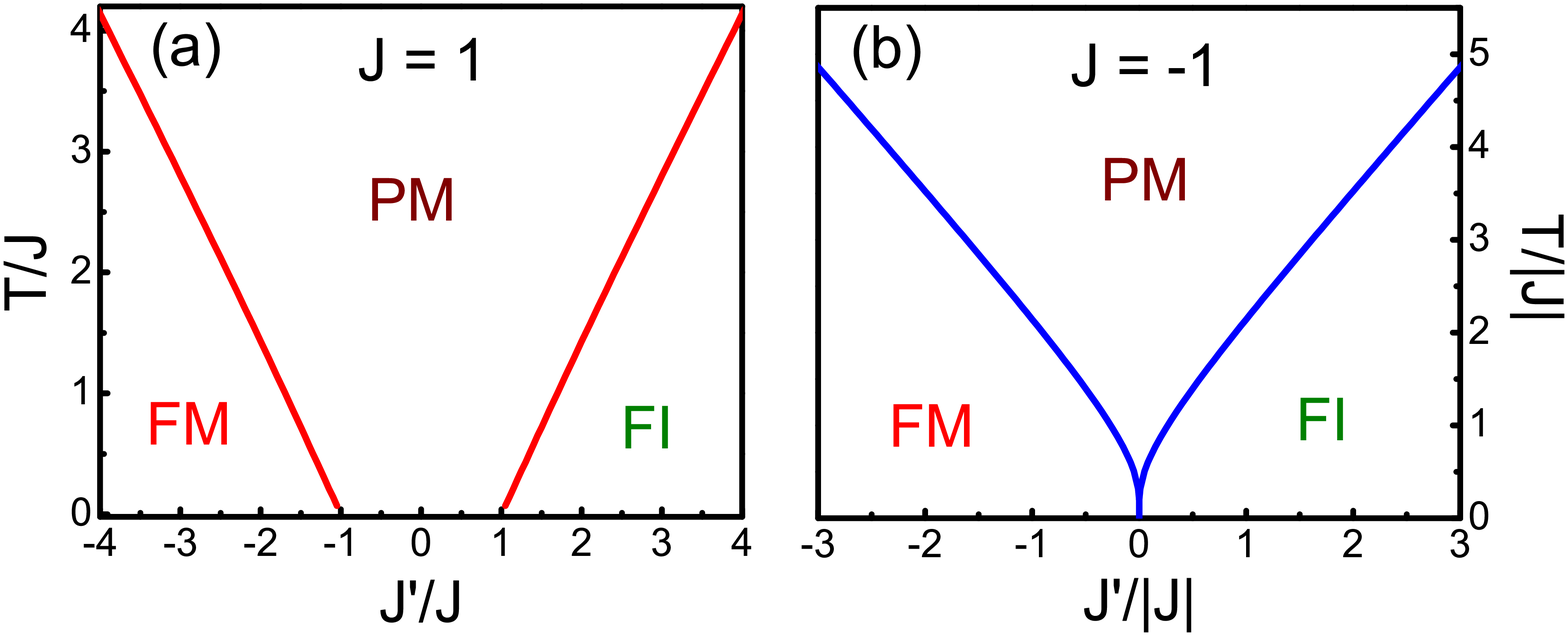}
\caption{(Color online) The zero-field phase diagrams for different
couplings: (a) $J>0$ and (b) $J<0$. The phase boundaries are
determined by Eq. (\ref{formula-Tc}). In the phase diagrams, FM means
ferromagnetic phase, FI represents ferrimagnetic phase, and PM
is paramagnetic phase.} \label{fig-phase-diag}
\end{figure}

Let us first utilize the exact mapping of 2D Ising model onto the
16-vertex model to present the exact solutions on the DK lattice.
Following Refs. \onlinecite{Diep,Diep1,Diep2}, for $h=0$, we can
write down the free energy per triangle in the thermodynamic limit as
\begin{eqnarray*}
f &=&-\frac{T}{16\pi ^{2}}\int_{-\pi }^{\pi }d\theta \int_{-\pi }^{\pi
}d\phi \,\ln [a+2b\cos (\theta ) \\
&+&2c\cos (\phi )+2d\cos (\theta -\phi )+2e\cos (\theta +\phi )],\text{ \ \
\ \ \ \ }(2)
\end{eqnarray*}%
where
\begin{eqnarray}
a &=&\omega _{1}^{2}+\omega _{2}^{2}+\omega _{3}^{2}+\omega _{4}^{2},  \notag
\\
b &=&\omega _{1}\omega _{3}-\omega _{2}\omega _{4},  \notag \\
c &=&\omega _{1}\omega _{4}-\omega _{2}\omega _{3},  \notag \\
d &=&\omega _{3}\omega _{4}-\omega _{7}\omega _{8},  \notag \\
e &=&\omega _{3}\omega _{4}-\omega _{5}\omega _{6},  \notag \\
\omega _{1} &=&2\exp (-2J/T)[1+\exp (2J/T)\cosh (2J^{\prime }/T)]^{2},
\notag \\
\omega _{2} &=&\omega _{1}-8\cosh (2J^{\prime }/T),  \notag \\
\omega _{3} &=&\omega _{4}=\omega _{5}=\omega _{6}=\exp (2J/T)\cosh
(4J^{\prime }/T)-\exp (2J/T),  \notag \\
\omega _{7} &=&\omega _{8}=\omega _{1}-4\exp (-2J/T),  \notag
\end{eqnarray}
where $k_B=1$ is presumed. It is straightforward to readily verify
that these $\omega$'s satisfy the free-fermion
conditions,\cite{Diep} showing that the Ising model on the DK
lattice is exactly solvable. The critical temperature $T_c$ at which
the phase transition takes place is determined by the following
equation
\begin{equation}
1 + 4 \exp(2 J/T_c) \cosh(2 J'/T_c) - \cosh(4 J'/T_c) = 0.
\label{formula-Tc}
\end{equation}

\begin{figure}[tbp]
\includegraphics[angle=0,width=0.9\linewidth]{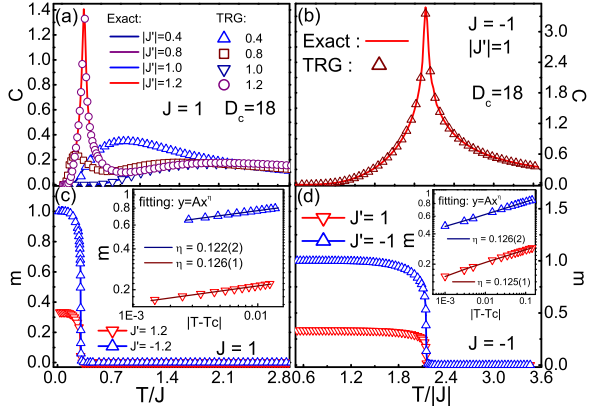}
\caption{(Color online) Temperature dependence of specific heat and
magnetization for different coupling ratio $J'/J$ at $h=0$. The TRG
results (open symbols) along with the exact solutions (solid and
dashed lines) are presented for (a) $J>0$ and (b) $J<0$. In (c) and
(d), the magnetization $m$ is plotted for $J>0$ and $J<0$,
respectively, where a magnetic order-disorder transition is clearly
seen. The insets illustrate the critical behaviors of $m$ near
$T_c$.} \label{fig-spec-heat}
\end{figure}

Two different cases are presented in Figs. \ref{fig-phase-diag} (a)
and (b). When $J>0$, the magnetic ordered phases appear when
$|J'/J|>1$, which can be recognized as a ferrimagnetic phase for
$J'/J>1$, and a ferromagnetic phase for $J'/J<-1$. This is obtained
by checking the magnitude of spontaneous magnetization [see
Fig.\ref{fig-spec-heat} (c) below]. The disordered paramagnetic
phase separates the two ordered phases, where the phase boundaries
are determined by Eq. (\ref{formula-Tc}). For a small $T_c$, Eq.
(\ref{formula-Tc}) can lead to a simple expression $T_c/J \approx
2(|J'/J|-1)/\ln4$, that gives the straight phase boundaries in Fig.
\ref{fig-phase-diag} (a). Notice that the paramagnetic phase exists
even at zero temperature owing to the frustration. When $J>0$ and
$|J'/J|\in[0,1]$, the spin surrounded by $J'$ couplings on $\alpha$
sublattice is free to flop up or down without costing energy.
Meanwhile, the ground-state spin configurations on $\beta$ and
$\gamma$ sublattices are highly degenerate. Hence, the total degeneracy is
$K=2^{N/3+N_c}$, with $N$ the number of total sites, and $N_c$ the
number of chains consisting of spins on $\beta$ and $\gamma$
sublattices (the horizontal lines in Fig. \ref{fig-DKL}). This
superdegenerate state at zero temperature connects continuously with
the paramagnetic phase at finite temperatures. No phase transition
appears, and the system is disordered at all temperatures. This
observation can also be manifested in Fig. \ref{fig-spec-heat} (a),
where no singularities exist in the specific heat for $|J'/J|=0.4,
0.8, 1.0$. However, for $|J'/J|>1$, as Fig. \ref{fig-phase-diag} (a)
indicates, there exists a ferromagnetic ($J'<0$) or ferrimagnetic
($J'>0$) state at low temperatures, and these ordered phases would
be destroyed through an order-disorder phase transition with
increasing temperature. Correspondingly, the specific heat for
$|J'/J|=1.2$ in Fig. \ref{fig-spec-heat} (a) shows a divergent peak,
which is a typical character of second-order phase transition. When
$J<0$ [Fig. \ref{fig-phase-diag} (b)], the situations are similar,
and the system is ordered (ferromagnetic or ferrimagnetic) at low
temperatures except for the case $J=-1, J'=0$, where the model is
degenerated into the decoupled one-dimensional Ising chains, which
has $T_c=0$ and thus is disordered at any finite temperature.

\section{TRG algorithm}

Exact solutions can offer us a reliable phase diagram of the model.
However, some other quantities such as the magnetization $m$ and
specific heat in nonzero magnetic fields cannot be obtained within
the above framework. Hence, we adopt the recently proposed TRG
numerical algorithm. The TRG method is first introduced to calculate
the 2D classical models,\cite{Levin,Chang} and then generalized to
study 2D quantum spin models.\cite{Jiang,Li,Gu,Chen} The principal
idea of TRG algorithm is to express the partition function (or the
expectation value of quantum operators) as a tensor network, and
then utilizes the coarse-graining and decimation procedures to
approximately obtain the results. TRG is an efficient method both
for classical and quantum spin models.

The first step is to replace each triangle on Kagom\'{e} lattice by
a tensor, as shown in Fig. \ref{fig-DKL} (a). The energy of each
triangle in an external magnetic field $h$ is $\varepsilon_\triangle
(s_1,s_2,s_3) = J' s_1 s_2+J' s_1 s_3+J s_2 s_3 -\frac{1}{2}
h(s_1+s_2+s_3)$. We introduce a three-order tensor
$T^{A/B}_{s_1,s_2,s_3} =
\exp(-\varepsilon_\triangle(s_1,s_2,s_3)/T)$, where A(B) means down
(up)-pointing triangle in Fig. \ref{fig-DKL} (a). These tensors form
a honeycomb lattice, and the partition function can be expressed as
\begin{eqnarray}
Z &  = & \sum_{s_1, s_2, s_3, ... =
-1,1}\exp\{-[\varepsilon_\triangle(s_1,s_2,s_3)+\varepsilon_\triangle(s_1,s_4,s_5)+...]/T\}
\nonumber \\
& = & \sum_{s_1, s_2, s_3, ... = -1,1}
T^A_{s_1,s_2,s_3} T^B_{s_1,s_4,s_5} ... = tTr(T^A T^B ...),
\label{formula-partition func}
\end{eqnarray}
where $tTr$ represents the tensor trace. Hence the problem of
solving the partition function of Ising model on a DK lattice is
equivalently transformed into a honeycomb tensor network problem,
which can be efficiently evaluated through the rewiring and
coarse-graining iterations (see the details in Ref.
\onlinecite{Jiang}). Upon obtaining the partition function $Z$,
other thermodynamic quantities can be evaluated straightforwardly.
Alternatively, we can also introduce some impurity tensors in the
tensor networks to achieve this goal. For example, in order to
calculate the magnetization $m$, an impurity tensor
$T_{s_1,s_2,s_3}^{Im}=(\frac{s_1+s_2+s_3}{3})
\exp[-1/T\varepsilon_{\triangle}(s_1,s_2,s_3)]$ can be introduced.
By replacing one tensor $T^{A/B}$ in Eq. (\ref{formula-partition
func}), we can get the magnetization per site
 \begin{equation}
m=\frac{tTr(T^{Im} T^B T^A ... )}{Z}. \label{formula-magnetization}
\end{equation}
In the following, the second scheme is adopted for evaluating the
thermodynamical quantities, such as the magnetization $m$, energy
per site $e$, etc.

In our calculations, the number of coarse-graining iterations is
generally taken as 20, i.e., the total sites of DK lattice under
investigation is $3^{22} \thickapprox 3 \times 10 ^{10}$, which is
close to the thermodynamic limit. In addition, the periodic boundary
conditions are adopted during the simulations. The initial bond
dimension D of tensor $T$ is chosen as 2 owing to the two states
(spin-up and -down) of the Ising spins. With the coarse-graining
procedure, the bond dimension will increase dramatically, and hence
we have to make a truncation and reserve a finite dimension $D_{c}$.
In our calculations, $D_{c}$ is taken as high as 18, and the
convergence with various $D_{c}$ has always been checked.

\begin{figure}[tbp]
\includegraphics[angle=0,width=0.7\linewidth]{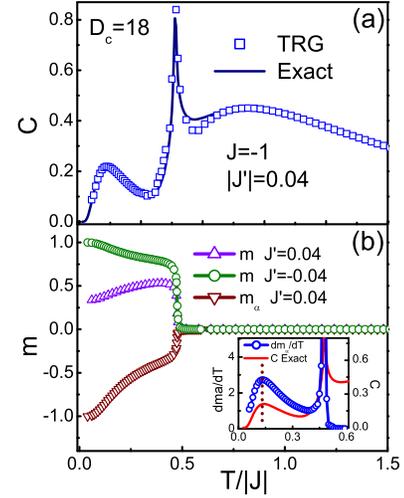}
\caption{(Color online) The specific heat and magnetization $m$ as
functions of temperature with $|J'/J|=0.04$ at $h=0$. (a) Specific
heat, where three peaks appear, one of which is divergent; (b)
Magnetization $m$ and sublattice magnetization $m_{\alpha}$, where
$|m_{\alpha}|$ decreases rapidly (but does not vanish) around the peak
position of the specific heat at low temperature, which is also
revealed as a local maximum of $dm_{\alpha}/d T$ in the inset.}
\label{fig-tri-peak-C}
\end{figure}

\section{Specific Heat and Phase Transitions}

When $h=0$, both exact solution and TRG method can be utilized to
evaluate the specific heat. In Figs. \ref{fig-spec-heat} (a) and
(b), the TRG results are plotted by symbols, while the exact results
by lines. Excellent agreement can be observed, except for the region
around the critical point where a divergent peak occurs. Another
character is that the specific heat at zero field is independent of
the sign of coupling $J'$, but is relevant to the magnitude of
$|J'|$. In Fig. \ref{fig-spec-heat} (a), when $|J'|$ is small, there
is only one round peak in the specific heat. By tuning $|J'|$ to
approach $J$ from below  (e.g. $|J'|/J = 0.8$), a new round peak
appears at low temperature, which disappears when $|J'|=J$ and the
system again exhibits a single round peak. These observations imply
that there exist no phase transitions when $|J'| \leq J$, which is
owing to the strong frustration, and is in accordance with the phase
diagram in Fig. \ref{fig-phase-diag} (a). Furthermore, if $|J'|$
exceeds $J$ [as $|J'/J|=1.2$ in the Fig. \ref{fig-spec-heat} (a)], a
divergent peak emerges, implying the occurrence of phase transition.
In Fig. \ref{fig-spec-heat} (b), a typical curve of specific heat
with $|J'/J|=1$ is shown. A divergent peak occurs at the transition
temperature $T_c \approx 2.14$, again in agreement with the exact
solution ($T_c=4/ \ln(3+2\sqrt3)$). Moreover, the specific heat is logarithmically divergent
at the critical point because of $2\exp(2 J/T_c)+\cosh(4 J'/T_c)-1
\neq 0$, as shown in Figs. \ref{fig-spec-heat} (a) and (b).

The phase transitions can also be verified by studying the order
parameter, i.e., the magnetization per site $m$ defined in Eq.
(\ref{formula-magnetization}). In Figs. \ref{fig-spec-heat} (c) and
(d), when $J=1, J'/J>1$ or $J=-1, J'>0$, $m=1/3$ at $T=0$ and
remains finite at small temperatures. This nonzero spontaneous
magnetization implies the existence of a ferrimagnetic phase; while
$J=1, J'/J<-1$ or $J=-1, J'<0$, the magnetization starts from $m=1$,
and the system is in a ferromagnetic phase when $T$ is smaller than
the critical temperature $T_c$. With increasing temperature, $m$
decreases steeply to zero in the vicinity of $T_c$, showing a
order-disorder phase transition happens. In addition, the critical
behavior of $m$ near $T_c$ has been investigated. In the insets of
Figs. \ref{fig-spec-heat} (c) and (d), in terms of $m \propto
(T_c-T)^\eta$, the fittings in different cases coincidentally give
$\eta \simeq 1/8$, which is the same as that of Ising model on a
square lattice.\cite{Yang} The phase transition occurring at $T_c$
probably falls into the universality class of 2D Ising models.

Another interesting case is shown in Fig. \ref{fig-tri-peak-C} (a),
where the temperature dependence of the specific heat is presented
for $J<0$ and $|J'| \ll |J|$. One may note that there are three
peaks, including two round peaks and a divergent one. Both exact
solutions and TRG method give the same results. In order to
investigate the origin of each peak, the TRG method is utilized to
calculate the magnetization $m$ and sublattice magnetization
$m_{\alpha}$. As shown in Fig. \ref{fig-tri-peak-C} (b), $m$ behaves
rather differently for $J'=0.04$ and $-0.04$, although the specific
heat coincides for both. When $J'=-0.04$, the ground state is
ferromagnetic, and $m$ decreases monotonously with increasing
temperature and vanishes sharply at critical temperature $T_c$. The
case with $J'=0.04$ is more interesting, where the system possesses
a ferrimagnetic ground state with $m=1/3$ at $T=0$. With increasing
temperature, $m$ first increases until the temperature is close to
the critical point $T_c$, and then goes down steeply to zero. In
order to understand this peculiar behavior, we have plotted the
sublattice magnetization $m_{\alpha}$ as a function of $T$ in Fig.
\ref{fig-tri-peak-C} (b). In the ferrimagnetic case, $m_{\alpha}$ is
aligned anti-parallel with the spins on the other two sublattices, and its magnitude
decreases rapidly with increasing temperature owing to the coupling
$J'$ weak. Hence, $m=(-|m_{\alpha}|+m_{\beta}+m_{\gamma})/3$ would
first increase until the temperature approaches to $T_c$, where
$m_{\alpha}$, $m_{\beta}$, and $m_{\gamma}$ disappear
simultaneously. Moreover, as the inset shows, the first-order
derivative $d m_{\alpha} /d T$ has a round peak at the temperature
$T_r$, which coincides with the low temperature peak position of
specific heat. Although $m_{\alpha}$ decreases rapidly around $T_r$,
it does not vanish. It should be pointed out that $T_r$ is not a
critical point, as the specific heat shows only a round peak and
never diverges at $T_r$.

\section{Magnetization and Susceptibility}

\subsection{Magnetization Plateaux and Ground State Phase Diagrams}

When the external magnetic field is switched on, the exact solution
in Sec. II no longer works. The TRG method, which has been verified
to be accurate and reliable in the previous sections, is utilized to
study the response of the system to an external magnetic field.

\begin{figure}[tbp]
\includegraphics[angle=0,width=1.0\linewidth]{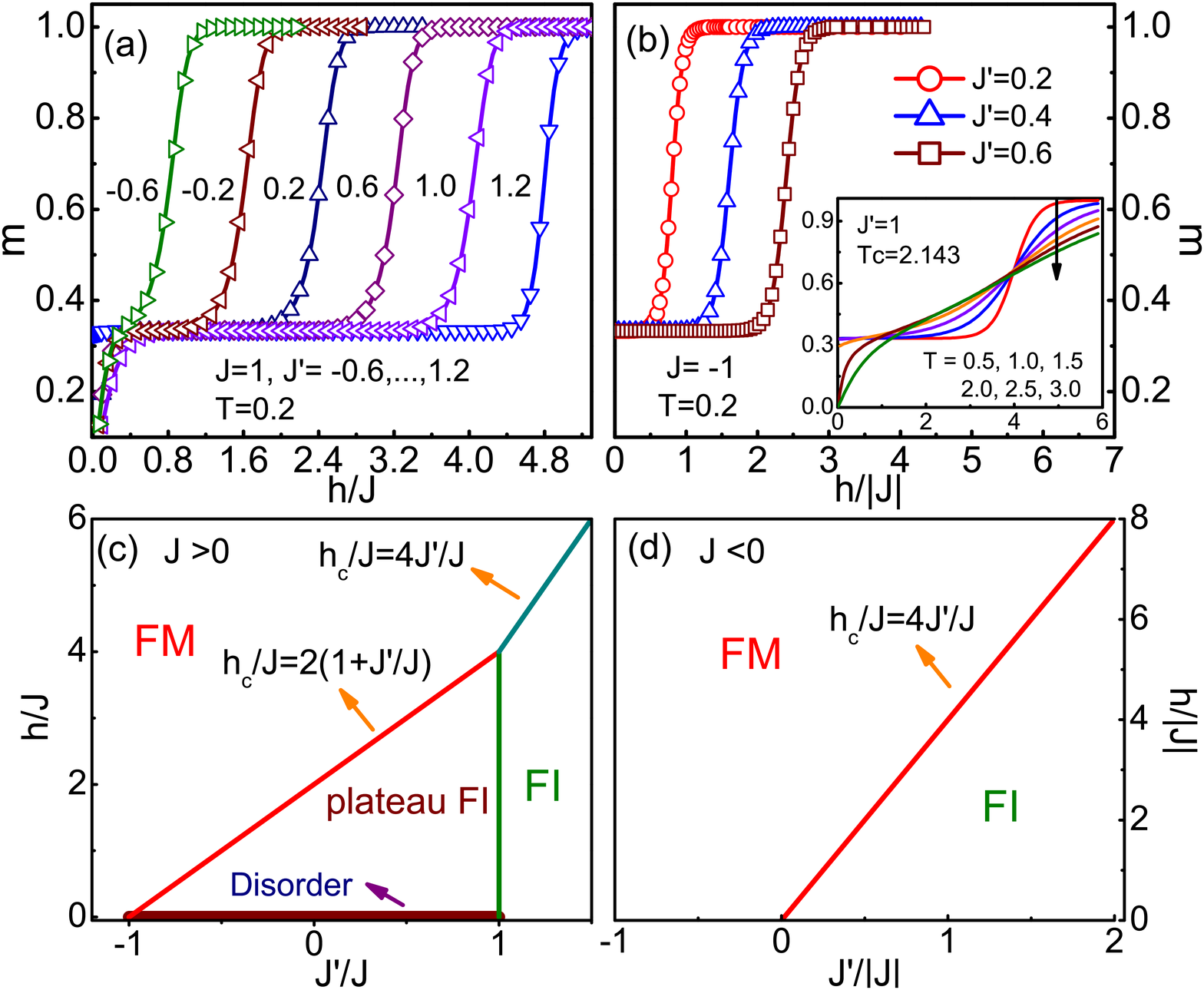}
\caption{(Color online) In (a) and (b), the magnetic curves for
different couplings for $J>0$ and $J<0$ are shown, respectively,
where $|T/J|=0.2$ and $D_c$=18. Inset in (b) presents the magnetic
curves with different temperatures below and above $T_c$. In (c) and
(d), the ground-state phase diagrams on $J'-h$ plane are presented.}
\label{fig-mag-curv}
\end{figure}

Let us first focus on the magnetization, where the $1/3$
magnetization plateaux are obtained, as shown in Fig.
\ref{fig-mag-curv}. When $J>0$ and $J'/J\in [-1,1]$ in Fig.
\ref{fig-mag-curv} (a), the system is in a paramagnetic phase at all
temperatures. An infinitesimal small magnetic field can polarize the
free spin on $\alpha$ sublattice at $T=0$, and hence drives the
ground state to a ferrimagnetic state with $m=1/3$ after a
magnetization jump, and the $1/3$ plateaux appear in the
magnetization curves. At finite but small temperatures, these
plateaux are still present. This field-induced $1/3$ plateau
ferrimagnetic phase is highly degenerate, and the degeneracy is $K =
2^{N_c}$, where $N_c$ is the number of independent spin chains in
the system. One of the degenerate spin configurations is shown in
Fig.\ref{fig-DKL} (b). When the field is larger than a critical
field $h_c$, the spins on $\beta$ and $\gamma$ sublattices align
parallel instead of antiparallel, and the system has the saturated
magnetization $m=1$, leading to a ferromagnetic spin configuration.
The energy difference per site between the polarized ferromagnetic
and the plateau ferrimagnetic state is $\delta e =
\frac{4}{3}(J'+J)$. The Zeeman energy $\delta e_z = -\frac{2}{3} h$
at the critical magnetic field $h_c$ has to compensate this energy
difference. $\delta e_z + \delta e = 0$ leads to $h_c=2(J+J')$, as
verified in Fig. \ref{fig-mag-curv} (a), where the critical field
$h_c$ increases with enhancing the coupling $J'$. In addition, there
exists a notable difference between the magnetization curves of
$J'=1.2$ and others in Fig. \ref{fig-mag-curv} (a). The former
starts from a nonzero spontaneous magnetization ($m=1/3$) owing to
its ferrimagnetic ground state instead of a paramagnetic one. The
ground state spin configuration for $J'>1$ is illustrated in Fig.
\ref{fig-DKL} (c), which is ferrimagnetically ordered. Considering
the spontaneously broken $Z_2$ symmetry, this ground state is no
longer degenerate. Hence, the width of $1/3$ plateau does not obey
the relation mentioned above, but has another relation in the
ferrimagnetic case to be discussed below.

\begin{figure}[tbp]
\includegraphics[angle=0,width=1.0\linewidth]{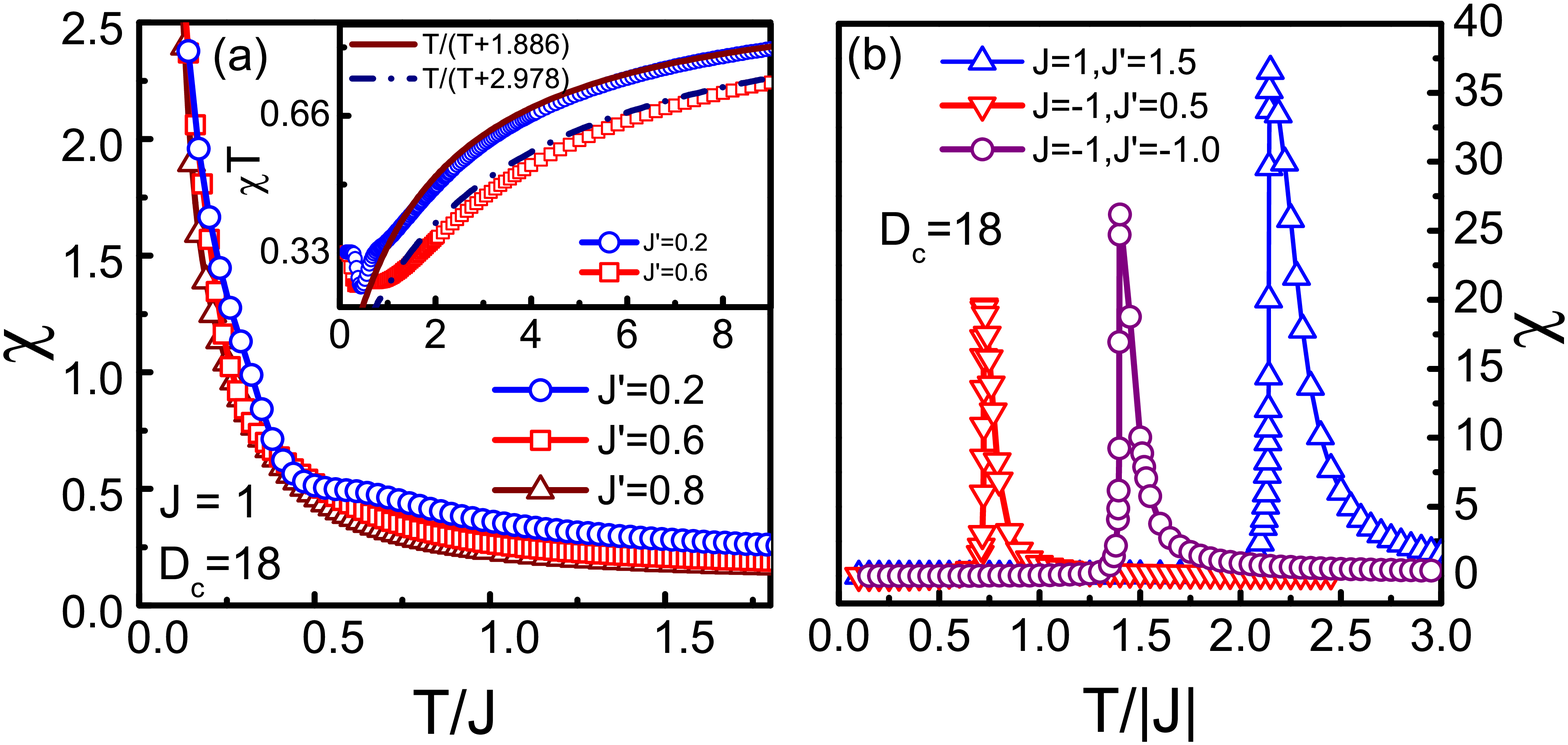}
\caption{(Color online) Temperature dependence of zero-field
susceptibility for different $J$ and $J'$. (a) $\chi$ diverges at
$T=0$ and $\chi T$ converges to 1/3 as the inset shows; (b) $\chi$
shows divergent peaks at the critical temperature $T_c$, where the
magnitude of $\chi$ with $J=J'=-1$ has been divided by two. The
susceptibility is calculated at $h=0.01$ and, the convergence with
various small fields has been checked.} \label{fig-suscep}
\end{figure}

For $J<0$, there exist ferromagnetic ($J'<0$) and ferrimagnetic
($J'>0$) ground states. The cases with $J'>0$ possess $m=1/3$
plateaux, as seen in Fig. \ref{fig-mag-curv} (b). Comparing with the
former case $J>0$, the width of $1/3$ plateaux has a different
relation with couplings $J$ and $J'$. By identifying $\delta e =
\frac{8}{3} J'$ and $\delta e_z = -\frac{2}{3} h$, $h_c = 4 J'$ is
obtained, that is independent of $J$. Here, the spin configuration
on $m=1/3$ plateau is illustrated in Fig. \ref{fig-DKL} (c). The
order parameter characterizing this phase is the spontaneous
magnetization $m|_{h=0}$, which implies the breaking of $Z_2$
symmetry. By increasing temperature, the $Z_2$ symmetry will
eventually recover above the critical temperature $T_c$, and the
spontaneous magnetization will vanish. As shown in the inset of Fig.
\ref{fig-mag-curv} (b), the magnetization at $T>T_c$ starts from
$m=0$, and the $1/3$ plateau is smeared and finally destroyed by
strong thermal fluctuations.

In order to look at the effects of external magnetic fields, the
phase diagrams at zero temperature are plotted in Figs.
\ref{fig-mag-curv}(c) and (d). For $J>0$, as shown in Fig.
\ref{fig-mag-curv} (c), there are four different phases including
ferromagnetic, ferrimagnetic, plateau ferrimagnetic, and disorder
phase that only exists in the $h=0$ line. It is worthwhile
emphasizing that although the magnetization in the $1/3$ plateau
ferrimagnetic phase has the same value as that in the ferrimagnetic
phase at $T=0$, they are quite different in nature. The former is
induced by a magnetic field and highly degenerate with the
degeneracy $K=2^{N_c}$, while the latter is an spontaneously ordered
phase with the $Z_2$ symmetry breaking. As indicated in Fig.
\ref{fig-mag-curv} (d), for $J<0$, only a ferrimagnetic phase and a
ferromagnetic phase exist. Here we would like to stress that the phase
transitions between these different phases only occur
at $T=0$, and the temperature would then
blur the transitions. In fact, in the presence of an external
magnetic field, there are no thermodynamic phase transitions at
finite temperatures, which will be discussed in Sec. VI.

\subsection{Susceptibility}

In order to understand the magnetic response of the present system
to an external magnetic field, the zero-field susceptibility $\chi$
is obtained by $\chi = [m(h)-m(h=0)]/h$ for a small magnetic field.
In Fig. \ref{fig-suscep} (a), where $J=1$ and $J'\in(0,1)$, the
ground state is disordered, and the spins on one ($\alpha$)
sublattice are free to flip up or down without an energy cost due to
the frustration effect. $\chi$ diverges, obeying Curie law, i.e.,
$\chi \propto 1/T$ as $T$ approaches zero. This result is validated
in the inset of Fig. \ref{fig-suscep} (a), where the $\chi T$ curves
converge to a constant $1/3$ at low temperatures, which is
independent of $J'$. The specific value $1/3$ can be attributed to
the free spins on one of three sublattices. On the other hand, in
the high temperature limit, $\chi$ decays with the Curie-Weiss law,
which can be fitted by $\chi T = \frac{T}{T+\theta}$ up to $T/J=100$
(Note that Fig. \ref{fig-suscep} shows only to $T/J\simeq9$). It is
straightforward to use the mean-field approximation to obtain the
Curie-Weiss temperature $\theta$ as $(8J'+4J)/3$, and $\theta=1.867$
and $2.933$ for $J'=0.2$ and $0.6$, respectively. The fittings in
the inset of Fig. \ref{fig-suscep} (a) agree with the mean-field
predictions, that further validates our TRG results. In Fig.
\ref{fig-suscep} (a), it is interesting to notice that there exists
a turning point at an intermediate temperature in the crossover
region, which separates the low $T$ Curie behavior and high $T$
Curie-Weiss behavior, as shown in the inset of Fig.
\ref{fig-suscep}(a). Quite differently, as seen in Fig.
\ref{fig-suscep} (b), when the ground state is ferrimagnetically or
ferromagnetically ordered, the susceptibility has a divergent peak
at $T_c$ where the magnetic ordering is destroyed by thermal
fluctuations. This again certifies the existence of magnetic
order-disorder phase transitions.

\begin{figure}[tbp]
\includegraphics[angle=0,width=1.0\linewidth]{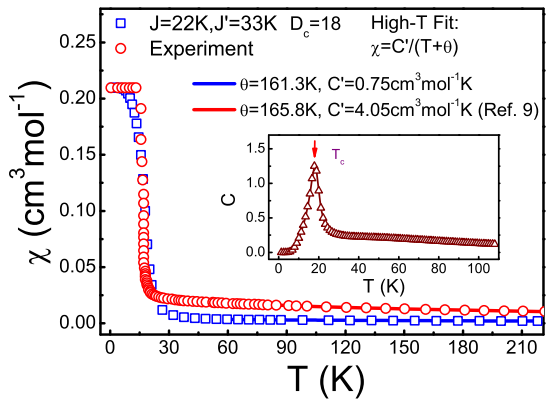}
\caption{(Color online) A comparison of TRG results to the
experiment, where the experimental data are taken from Ref.
\onlinecite{Gao}. The TRG result is calculated at a small field
$h/J=0.05$. The high temperature fittings with the Curie-Weiss law
to both experimental and TRG results are also shown. The inset
predicts a divergent peak in the specific heat around $T=20$K.}
\label{fig-fit exp}
\end{figure}

\subsection{Comparison to Experiments}

The complex Co(N$_3$)$_2$(bpg)$\cdot$ DMF$_{4/3}$ reported in Ref.
\onlinecite{Gao} is a molecular magnetic material, in which the
Co$^{+2}$ ions form a distorted Kagom\'{e} layer. Experimentally,
the susceptibility does not go to zero as $T$ approaches zero (see
Fig. \ref{fig-fit exp}), which is unusual for an isotropic
Heisenberg antiferromagnetic system. In fact, the Co$^{+2}$ ions are
believed to have effective spin-1/2 when $T\leq$ 20K, with
anisotropic Lande $g$ factors ( $g_\parallel \neq 0$, $g_\perp
\approx 0$), which implies that in this compound the Ising-type
couplings may be dominant between Co$^{+2}$
ions.\cite{Carlin,xyWang} Here, we try to use our TRG results to fit
the experimental data of susceptibility (especially for the low $T$
region) for this complex. To be consistent with the experimental
convention, the definition of susceptibility $\chi = m(h)/h$ is
adopted. As shown in Fig. \ref{fig-fit exp}, $\chi$ decreases
steeply around the transition temperature and, one may see that the
fittings agree rather well with the experimental data at low
temperatures. The exchange coupling constants for this compound are
estimated through the fittings as $J=22K$ and $J'=33K$. According to
our study on the Ising DK lattice with the parameters $J>0$ and
$J'/J=1.5$, the system has a ferrimagnetic phase at low
temperatures. It is thus not difficult to understand why the low
temperature susceptibility goes to a finite value instead of zero
for this compound. At high temperatures, by fitting the TRG results
with the Curie-Weiss law $\chi = C'/(T+\theta)$, we find that the
Curie-Weiss temperature $\theta \thickapprox 161.3K$, which agrees
well with that of experimental estimation ($\theta \thickapprox
165.8K$, see online supporting material of Ref. \onlinecite{Gao}).
Besides, we find that the ratio of the experimental susceptibility
to the result from the Ising model equals a constant $R \approx 5.4$
in the high temperature limit. This constant ratio may be ascribed to
the fact that in the material at $T >$20K the effective spin of
Co$^{+2}$ ions may no longer be 1/2 and also, the other interactions
such as XY couplings may intervene, giving rise to that the Ising
model is insufficient to describe the behaviors of this complex.
Surely, more experimental results towards this direction are needed.
In addition, we have calculated the specific heat based on the Ising
model with the couplings given above, and found that a divergent
peak exists around $T=20$K, as depicted in the inset of Fig.
\ref{fig-fit exp}, suggesting that this compound may undergo a phase
transition at low temperature. Experimental studies on the specific
heat and other quantities for this compound will be carried out in
near future.

\begin{figure}[tbp]
\includegraphics[angle=0,width=1.0\linewidth]{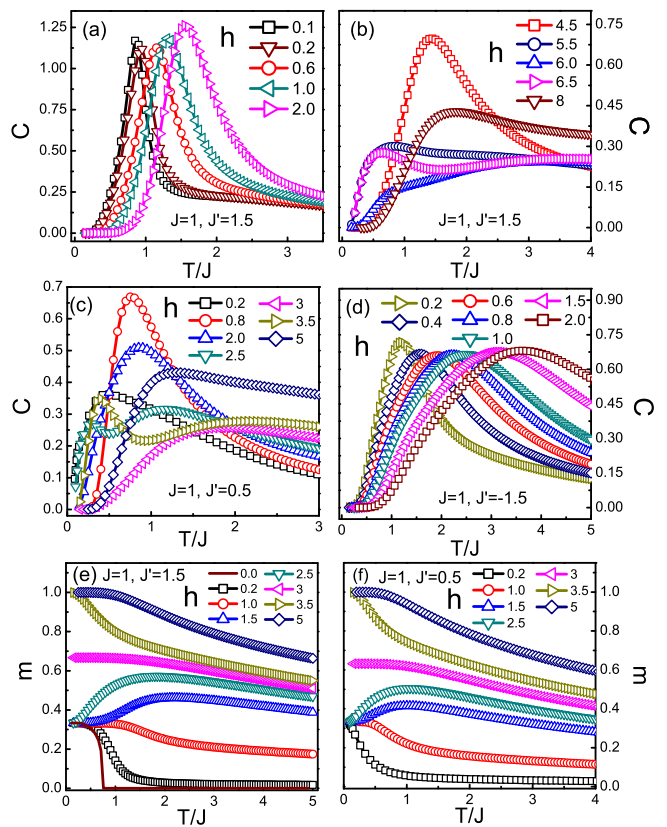}
\caption{(Color online) The specific heat $C$ and the magnetization
$m$ as functions of temperature in different magnetic fields. (a),
(b) and (e) illustrate the ferrimagnetic case $J>0, J'/J=1.5$, (c)
and (f) are for the paramagnetic case $J>0, J'/J=0.5$, and (d)
depicts the ferromagnetic case $J>0, J'/J=-1.5$.}
\label{fig-spec
field}
\end{figure}

\section{Specific Heat and magnetization in a Magnetic Field}

Next, we will study the effect of an external magnetic field on the
specific heat. Three typical cases will be studied: ferrimagnetic
($J>0, J'/J=1.5$), paramagnetic ($J>0, J'/J=-0.5$), and
ferromagnetic ($J>0, J'/J=-1.5$) cases.

In Fig. \ref{fig-spec field}, the specific heat in the presence of
an external field for the ferrimagnetic case ($J>0$, $J'/J>1$) is
shown. Fig. \ref{fig-spec field}(a) shows at small fields with $h
\leq 2J$, the peak of specific heat moves towards high temperatures,
and its height firstly decreases, and then increases with enhancing
the field until it approaches the spin flop critical field $h_c$,
which polarizes all spins. In Fig. \ref{fig-spec field}(b), when the
field keeps increasing, the peak of specific heat becomes dulled,
and then splits into double peaks (except for the point $h=h_c$),
which can be viewed as a field-induced splitting. Similar phenomena
have also been observed in other Ising and Heisenberg spin
systems.\cite{Gong,Li2} The double peak scenario will eventually be
spoiled by further increasing $h$. When $h \gg h_c$, the specific
heat will again be single-peaked. It is notable that the divergent
peaks at zero field disappear immediately when the field is switched
on, which means that the phase transitions are absent and the system
remains in the ferrimagnetic phase at all temperatures. When $h=0$,
the ferrimagnetic ordered phase spontaneously breaks the $Z_2$
symmetry contained in the Hamiltonian (see Eq.
\ref{formular-Hamiltonian}), and possesses a nonzero order
parameter. When $T>T_c$, the thermal fluctuations will destroy the
magnetic order, while $Z_2$ symmetry will be recovered, and $m$
vanishes immediately [the solid line in Fig. \ref{fig-spec field}
(e)]. However, the external field explicitly breaks the $Z_2$
symmetry in the Hamiltonian, and $m$ is nonzero even at high
temperature $T>T_c$ [the symbol lines in Fig. \ref{fig-spec field}
(e)]. Therefore, no phase transition occurs in the presence of a
magnetic field. In Fig. \ref{fig-spec field}(e), according to the
magnetization $m$ at zero temperature, the curves can be classified
into three classes. When $h<h_c$, the curves start from $m=1/3$;
while $h>h_c$, the spins are polarized and $m=1$ at $T=0$; when
$h=h_c$, the case is of a little subtlety, where $m$ equals to the
statistical mean value $2/3$ at zero temperature.

In Figs. \ref{fig-spec field} (c) and (f), the paramagnetic case
$J>0, J'/J=0.5$ is studied. The field will firstly promote the peak
height of the specific heat, and moves the peak to the high
temperature side. When $h$ is close to the critical field, the
specific heat will again be dulled, where the height is decreasing,
and the peak splits into two sub peaks except at the point $h=h_c$.
When the field $h\gg h_c$ a single peak of the specific heat recurs.
The $m-T$ curves in Fig. \ref{fig-spec field} (f) are quite similar
to those in Fig. \ref{fig-spec field} (e) and can be classified
analogously. At last, the ferromagnetic case $J>0, J'/J<-1$ is shown
in Fig. \ref{fig-spec field} (d). The divergent peak for the
ferromagnetic-paramagnetic phase transition disappears owing to the
same reason in the ferrimagnetic case as mentioned above. The
specific heat reveals a round peak, which moves towards the high
temperature side. By continuously enhancing the field, the height of
the peak decreases down firstly and then goes slowly up.

Besides, we have also studied other situations with different
couplings, and found that they can be ascribed into the above three
classes. For instance, other ferrimagnetic cases with $J<0,J'>0$ and
ferromagnetic cases with all ferromagnetic couplings ($J, J'<0$)
behave similarly with those presented in Fig. \ref{fig-spec field}.

\section{Summary and discussion}

In this article, we have systematically studied the thermodynamics
and magnetic properties of Ising model on a DK lattice by exact
solutions and the TRG numerical method. It is shown that the phase
diagrams are composed of three phases including ferromagnetic,
ferrimagnetic, and paramagnetic phases. Phase transitions between
them are identified by studying the specific heat and magnetization.
The critical exponent $\eta$ of $m$ near $T_c$ is determined as
$1/8$, which appears to fall into the universality of the 2D Ising
models. The TRG results of zero-field specific heat agree very well
with the exact solutions, showing that TRG is an efficient and
accurate tool in dealing with 2D Ising models. The TRG method is
also utilized to study the properties in the presence of a magnetic
field. In the magnetization curves, $1/3$ plateaux at low $T$ are
identified and, the relations of the plateau width with coupling
constants $J, J'$ are obtained. In addition, the zero temperature $J'-h$ phase
diagrams are presented to clarify the various ground state phases
in external magnetic field. The zero-field susceptibility
$\chi$ of the paramagnetic case ($J>0, |J'/J|\leq 1$) is found to
obey Curie law at low $T$ and Curie-Weiss law at high $T$. While in
the ferrimagnetic or ferromagnetic case, the divergent peak of
$\chi$ is found at the critical temperature. Moreover, the specific
heat under different magnetic fields is also investigated. It is
uncovered that the phase transitions are absent immediately when a
magnetic field is switched on, and the field-induced peak splitting
of the specific heat is recognized when $h$ is close to the critical
field. We have also fitted the experimental data of susceptibility
of the complex Co(N$_3$)$_2$(bpg)$\cdot$ DMF$_{4/3}$ with the TRG
results, and obtained the couplings $J=22$K and $J'=33$K. Based on
TRG calculations, a ferrimagnetic-paramagnetic phase transition is
expected to occur at about $T=20$K in this complex, which will be
studied in future. The present study offers a systematic
understanding for physical properties of the 2D Ising model on the
DK lattice, and will be useful for analyzing future experimental
observations in related magnetic materials with DK lattices.

\acknowledgements
We are indebted to Z. Y. Chen, Y. T. Hu, X. L. Sheng, Z. C. Wang, X. Y. Wang,
B. Xi, Q. B. Yan, F. Ye, and Q. R. Zheng for helpful discussions. This work is
supported in part by the NSFC (Grants No. 10625419, No. 10934008,
No. 90922033) and the Chinese Academy of Sciences.

\end{document}